\documentclass{llncs}
\pdfoutput=1 

\usepackage[ruled]{algorithm2e}
\usepackage{graphicx}
\usepackage{amssymb}
\usepackage{url}
\usepackage{siunitx}

\begin{document}
\mainmatter

\title{Reaching Consensus Among Mobile Agents: A Distributed Protocol for the Detection of Social Situations}


\author{
Daniel Raumer, Christoph Fuchs, and Georg Groh
}
\institute{ \email{\{raumer\textbar fuchsc\textbar grohg\}@in.tum.de}
\\TU M\"unchen, Department of Computer Science,  Boltzmannstr. 3, D-85748 Garching, Germany}

\maketitle
\begin{abstract}
Physical social encounters are governed by a set of socio-psychological behavioral rules with a high degree of uniform validity. Past research has shown how these rules or the resulting properties of the encounters (e.g. the geometry of interaction) can be used for algorithmic detection of social interaction. 
In this paper, we present a distributed protocol to gain a common understanding of the existing social situations among  agents. 

Our approach allows a group of agents to combine their subjective assessment of an ongoing social situation. 
Based on perceived social cues obtained from raw data signals, they reach a consensus about the existence, parameters, and participants of a social situation.
We evaluate our protocol using two real-world datasets with social interaction information and additional synthetic data generated by our social-aware mobility model.
\end{abstract}

%
%

\section{Introduction}
Mobile devices can be used to algorithmically detect 
social situations by combining and analyzing sensor information (e.g. \cite{Groh2011,Groh2010,Matic12,Altshuler2012b,Pan2011,Akbas2013}). To allow new and promising applications it is essential to detect social situations with low latency (i.e. during the social encounter). Furthermore, the detection on distributed devices and the agreement among these makes it viable for applications on personal mobile devices. 
Therefore Groh et al.~\cite{Groh2011} presented an approach based on Subjective Logic (SL) to share opinions regarding the existence of social situations among agents on different levels of abstraction (raw low-level sensor data, ``sub-symbolic'' probabilistic models and ``symbolic'' social situation models). We enhance this approach and designed a distributed algorithm to assess the boundaries of social situations among agents.

In section\,\ref{sec:PreviousWork}, we summarize the required foundations of social situations, briefly describe related work, and provide a short overview of Subjective Logic. In section\,\ref{sec:OurWork}, we present our approach to reach consensus regarding the existing social situations within a group of agents. In section\,\ref{sec:evaluation}, we describe our evaluation of the algorithm. 
We conclude with a summary in section\,\ref{sec:Conclusion}.

\section{Previous and Related Work}
\label{sec:PreviousWork}

\subsection{Social Situations}
Adam Kendon's F-formations and the concept of ``o-spaces'' are widely used to analyze social interaction patterns~\cite{f-formation2,Marshall2011}. The term {``}o-space'' refers to the circular area which is formed by a group of socially interacting people (standing in a circle, shoulder to shoulder).  F-formations describe the spatial orientation of interacting individuals.
Mathematically, a social situation $S$ may be defined \cite{Groh2011} as a tuple
$(P,\widetilde{X})$. $P$ is a set of (unique identifiers for) socially interacting individuals who are fully mutually aware of their interaction and $\widetilde{X}$ is a spatio-temporal
reference.

\begin{figure}
\begin{center}
\includegraphics[scale=0.33]{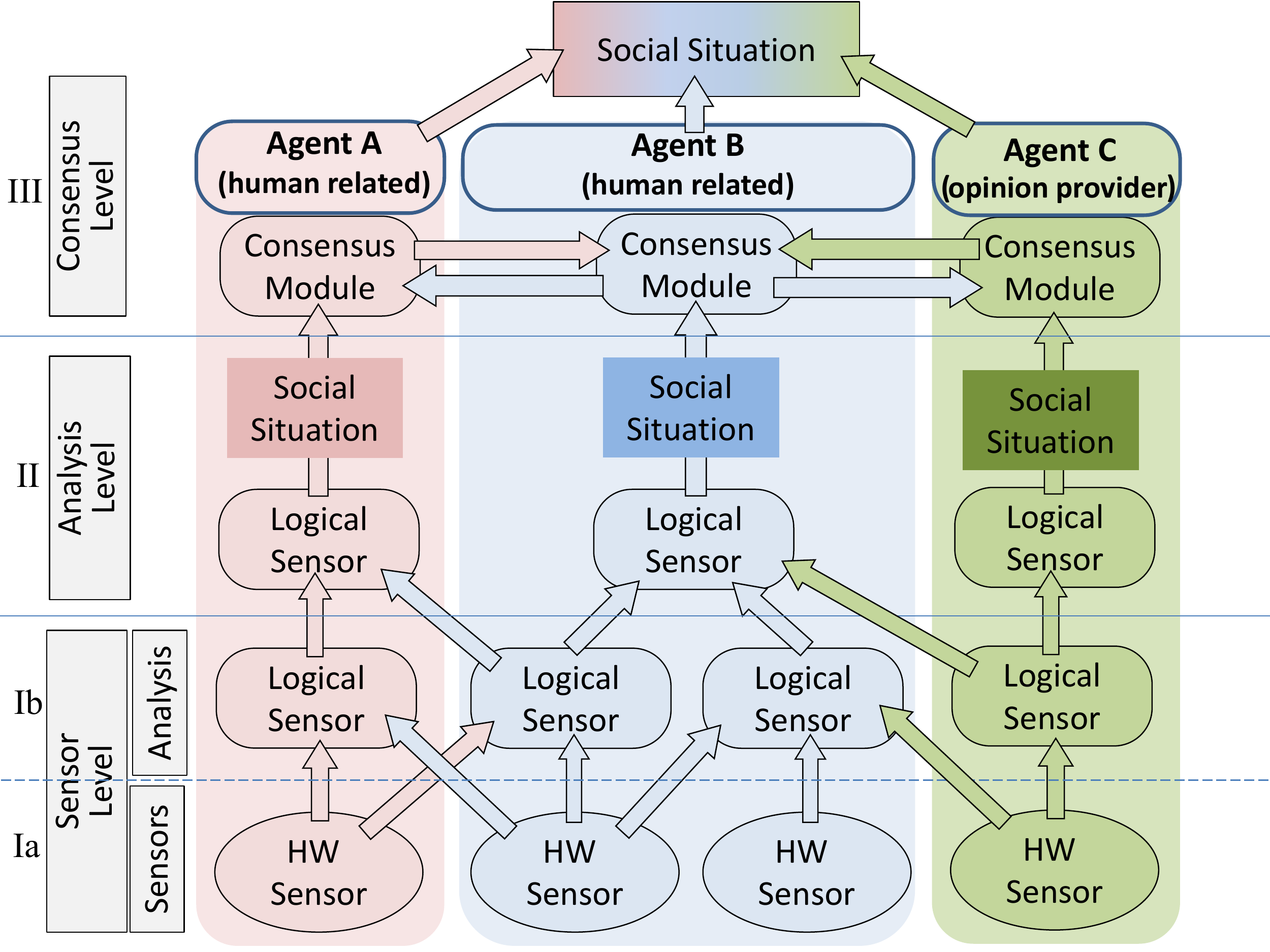}
\end{center}

\caption{\label{fig:architecture}Exemplary setting for distributed social situation detection}
\end{figure}

Groh et al. discussed \cite{Groh2011} how evidence for social situations from
several sensor sources can be exchanged and aggregated to algorithmically assess the existence of a social situation.
Fig.\,\ref{fig:architecture} shows the basic elements of Groh et al.'s architecture for ``social life networks''~\cite{Gupta13}.
It illustrates a setting of two agents, each directly connected to a human individual and an infrastructure related agent that is only entitled to provide opinions (see discussion in section\,\ref{sub:Disaggregating-agents-and}). 
Recent studies \cite{Matic12} demonstrated how smartphones can be used as agents for distributed social sensing.

The architecture has three conceptual layers: the
sensor level consists of a set of hardware sensors
(e.g. microphones, gyroscopes) providing raw data from the agent's
environment. Raw data is aggregated using sensor fusion to
either enhance quality of measurement (e.g. by using competitive
fusion for dependent opinions, i.e. measuring the same environmental
phenomenon with different means) or increase the span of information
(e.g. by including additional information not covered by the other
sensor) \cite[p. 32ff]{Brooks1997}. Some sensors may be better
characterized as an abstraction of a group of hardware sensors (and
therefore are part of layer Ib) while others include higher levels
of analysis (and thus belong to layer II). Layer II sensors are expected to output
the agent's subjective assessment of the user's current social situation.
Since a social situation requires full mutual awareness of the social situation's
existence among the proposed participants, layer III consists of a consensus module allowing the agents
to gain a common understanding of their social situation. Groh et al.~\cite{Groh2011,Groh2011b}
evaluated the layers I and II but did not consider a simulation of layer III. In section\,\ref{sec:OurWork}, we propose a concrete algorithm to establish consensus on social situations. 

\subsection{Agreeing on Social Situations}
\label{sec:ss}

\paragraph{Consensus finding}
In agent networks the term ``consensus'' refers to ``reach{[}ing{]}
an agreement regarding a certain quantity of interest that depends
on the state of all agents'' \cite{Olfati2007}. Consensus problems
have a long history reaching back to 1960s \cite{DeGroot1972}. Applications include flocking / swarming, sensor fusion, random networks, synchronization of coupled oscillators, etc. 
A  comprehensive overview is provided by Olfati et al.~\cite{Olfati2007}. With upcoming car-to-car
communication, certain aspects have been revisited
with high economic interest. 
Well-known protocols to find consensus include simple
quorum-based approaches (the option receiving the majority of votes is
seen as consensus), classical leader-follower architectures (all
other agents adopt the opinion of the leader), iterative approaches
like the one presented by DeGroot \cite{DeGroot1972} where each agent
revises its position after seeing the others' opinion, or the Monotonic
Concession Protocol \cite{Wooldrige2002} which allows finding an
optimal consent for two agents with different utility functions. While consensus
problems appear in various fields, to the best of our knowledge,
consensus regarding existing social situations in social ad hoc networks has not
been addressed. 
\paragraph{Clustering of ad hoc networks}
Traditional dis\-tri\-bu\-ted clustering algorithms often present a solution
to locally cluster data which is distributed across several nodes. Therefore, data gets clustered locally and results
are aggregated afterwards. These approaches are not applicable for our scenario as we wish to cluster the set of agents itself. 

Clustering of ad hoc networks corresponds roughly to our problem statement.
Clustering in ad hoc networks is mostly driven by designing sophisticated
routing mechanisms~\cite{Heinzelman2000,Younis2004} --
the fundamental idea is that data dissemination in networks is
faster and causes less effort when a multi-hop clustered topology is used.
Each cluster has a cluster head coordinating the cluster. Different
methods to nominate a cluster head are discussed by Chatterjee et al. in \cite{Chatterjee2002}.
Most clustering approaches distinguish between two phases: a set up
phase where the agent set is clustered and a maintenance phase to
adopt the clustering to the changing topology of the network (e.g.
due to moving or breaking nodes). For further details, please consider
\cite{Abbasi2007,Agarwal2007,Shayeb2011,Yu2005}
as an exemplary list of surveys on this topic.


\subsection{Subjective Logic}
\label{sec:subjectivelogic}
Subjective Logic \cite{Josang2001,Josang2007b} is an enhancement of the Dempster-Shafer theory of evidence~\cite{Shafer1971}.
Its main goal is to express uncertainty and subjectivity of statements made by agents. 
Assuming an atomic world state space $\Theta=\{x_{1},x_{2},...,x_{n}\}$, 
in {\em first order logic}, an agent's assessment of the world's state is either true or false. Subjectivity can be modeled by including statements of several agents. However, it is hard to model uncertainty.
{\em Classic probability theory} allows to model uncertainty but lacks subjectivity. Thus, it is difficult to combine statements from an a-priori unknown set of agents.

A Belief Mass Assignment (BMA) $m:2^{\Theta}\mapsto[0,1]$ assigns
a belief mass $m(x)$ to a subset $x\subseteq2^{\Theta}$ so that
$m(\emptyset)=0$, $m(x)\geq0$ and $\sum_{x\in2^{\Theta}}m(x)=1$.
An agent's belief in the statement ``the world
is in state $\eta$'' is expressed as $b(\eta)=\sum_{\eta'\subseteq\eta}m(\eta')$.

A simplified Dirichlet BMA (DBMA) assigns belief mass only to atomic
states and $\Theta$ as a whole, i.e. $b(x)\neq0\Rightarrow(x\in\Theta)\vee(x=\Theta)$.
Thus, an agent has a corresponding belief and an uncertainty
(expressed by assigning belief to $\Theta$). The base rate $a:\Theta\mapsto\text{[0,1]}$
can be interpreted as an assignment of a-priori probabilities for each
state. A multinomial opinion of an agent $A$ about a state $x$ is
defined as a tuple $\omega_{x}^{A}=(b,u,a)$ based on a DBMA. The
expected value of a state $x$ can be calculated by using the probability
expectation function $p(x)=b(x)+a(x)u$. It is comparable
to a posterior probability. 
A binary opinion about a state $x$,
denoted as $\omega_{x}=(b=b(x),d=b(\overline{x}),a=a(x))$, is a multinomial
opinion over a binary set $\Theta=\{x,\overline{x}\}$. The belief
in $\overline{x}$ can therefore be interpreted as disbelief in $x$
(written as $d(x)$).

The cumulative fusion operator $\oplus$ combines independent opinions (e.g. observations covering disjoint time intervals), whereas the averaging fusion $\underline{\oplus}$ operator is used to combine dependent opinions (e.g. observations covering the same time interval).
For a more detailed explanation, please refer to \cite{Josang2007b,Groh2011}.

\section{A Dis\-tri\-bu\-ted Protocol for Consensus on Social Situations}
\label{sec:OurWork}
%

\subsection{Basic Concept}
Reaching agreement about social situations among agents corresponds
to agreeing on clustering of nodes of a network only by means of local knowledge. 
Therefore, the terms {``}cluster'' and {``}social situation'' are used synonymously in this section (the same applies to {``}agent'' and {``}node'').
Since a multi-phase protocol including a set up and maintenance phase (as often used to cluster ad hoc networks)
is not practicable for a long running and constantly changing stream
of social situations we decided to use a protocol involving an arbiter.
Clusters are formed using an incremental process adding agent
after agent to a cluster. Agents who want to join
a cluster trigger a new agreement process where both parties
(the joining agent and the group of agents in the existing social
situation) need to agree. This procedure corresponds to the emergence
of real-life human social situations.
As in real life, social situations can merge.

Each social situation with $n$ members consists of one cluster head and $(n-1)$ members
(with $n\geq1$). The algorithm starts with every agent $i$ being
in a social situation with itself and therefore being the cluster head of its unary cluster $c_{i}=\{i\}$ ($i$ represents an unique identifier for this agent).

Each cluster head $i$ can decide to ask another agent $j$ to agree
on being in a joint social situation. If $j$ agrees, $i$ changes its role
to cluster member and $j$ becomes new cluster head of the joint social
situation (which is then referred to as $c_{j}$ since $j$ is the
new cluster head). Cluster members remain passive, requests received
to initiate social situations are forwarded to their respective cluster head.
Cluster members periodically broadcast opinions about ongoing social
interactions between pairs of nodes to all other nodes within communication
range. This is done using {``}member messages'' $m_{cm}$. Each cluster head
collects this information, periodically combines it with its own opinions (using the SL formalism discussed in section\,\ref{sec:subjectivelogic}),
and broadcasts the resulting group structure of its cluster in a {``}cluster head
message'' $m_{ch}$ to all members of its cluster (and thus to all
agents belonging to the cluster head's social situation). 

If a member $i$ leaves a social situation, the cluster head will reflect
this by not listing $i$ in the member list of the next cluster head message $m_{ch}$. The affiliation of agent $i$ to a social situation is decided by the cluster head based on the group's opinion. If the
cluster head itself leaves the social situation, the cluster breaks: A non-existent
cluster head causes an absence of the periodically sent $m_{ch}$ message
for a given period of time. This is a signal for the affected member
agents that the cluster is broken, i.e. the social situation has ended.
In both cases an agent not associated to any cluster forms its own cluster
again and changes its role back to cluster head. 

\subsection{Message Types\label{sub:Message-types}}


\textbf{Member messages $m_{cm}$} are sent by cluster members for two reasons: to inform
the environment (i.e. the nodes within communication range) about
the agent's assessment of the existing social situations and to inform
the cluster head that the sending member is still within reach (keep
alive function). The message contains one or more opinions about agents
being in a (pairwise) social situation, i.e. $m_{cm}$ sent by agent
$o$ in cluster $c_{p}$ consists of $o$'s identifier, the identifier
of $o$'s current cluster head $p$, and an arbitrary number of tuples
$ $ $(i,j,\omega_{i,j}^{o})$ with $\omega_{i,j}^{o}$ being $o$'s
SL opinion about the existence of a social situation
between agents $i$ and $j$. In a very trivial case with only one
single opinion as part of the message, $m_{cm}$ is a tuple $(o,p,(i,j,\omega_{i,j}^{o}))$.
The identifier of the current cluster head is included to allow the cluster head
to recognize if agents make false assumptions about their cluster head. Usually an
agent sends information concerning all agents within a socially relevant
distance. 
A distance of approx. $10m$ would be ideal \cite{Vinciarelli2009}.
If correspondingly accurate absolute device localization is not available, approximate relative positioning e.g. with the help of Wifi or Bluetooth is also possible \cite{Matic12}.
The message is sent with an agreed minimum frequency which can pragmatically be adjusted to the typical dynamic range of human social situations (e.g. 1Hz) to allow the cluster head
to recognize when a member node has left the cluster.

%
\textbf{Cluster head messages $m_{ch}$} are sent by cluster heads. As for the member
message discussed above, the purpose is twofold: 1) Propagate the
knowledge about the social situation that is managed by the
emitting cluster head and 2) inform the members of the cluster that
the cluster is still alive. The message can be described as a tuple
$m_{ch}=(p,c)$ with $p$ being the unique identifier of the cluster head
and $c$ being the set of identifiers of the cluster's members. The
set $c$ is built based on the preceding agreement process (i.e. the aggregation of the member messages $m_{cm}$ and the cluster head's opinion). Since
the underlying network is dynamic due to agent movement, the message
needs to be sent periodically by the cluster head with a fixed frequency
$f_{min}$ to indicate that the cluster still exists. $f_{min}$ can also be adjusted to typical human social situation dynamics. Member agents recognize that they
are out of communication range or left the cluster when they do not
receive a $m_{ch}$ message for a time $t>T=1/f_{min}$.

%
\textbf{Request messages $m_{req}$} are sent by cluster heads only. Their purpose is to
request agents to agree on being in a joint social situation.
The message can be described as $m_{req}=(p,c)\mbox{}$ with $p$
being the unique identifier of the cluster head and $c$ the set of
identifiers of the other agents in the respective cluster. Pragmatic agents only request
social situations if the corresponding aggregated belief reaches an appropriate threshold
and if the other agent is within a socially relevant distance to reduce energy consumption.

%
\textbf{Response messages $m_{res}$} are sent as a reaction to a request message $m_{req}$.
They inform the requesting party whether the request
was accepted or not. If the request was accepted, the {\em requesting node}
changes its status from cluster head to cluster member. The new cluster head
of the extended social situation is the {\em accepting agent}. If the request
gets declined, two cases are distinguished:
1) If the requested agent declines
the request since it is not allowed to manage the social situation (i.e. it is only a member and not the cluster head),
the negative reply includes the identifier of the cluster head of the
agent's social situation and all of its members. In
this case, the requesting node evaluates whether a social situation
with the complete cluster of the requested agent is feasible. If this
is the case, the request will be sent again to the cluster head of
the respective social situation. 2) If the requested agent declines
the request because it does not believe to be in a social situation with the requester, the requesting agent
stores this information for a certain period of time to avoid sending the request again.
 
\subsection{Agent Programs}
\label{sub:Agent-programs}

Algorithm\,\ref{alg1} shows the main process running on all
agents. Each agent starts in a separate cluster as a cluster head.
The agent periodically sends either a $m_{ch}$ message in case it is cluster head
or a $m_{cm}$ message if it is member of a cluster using a fixed time interval $T$.
%
\begin{algorithm}[htb]
$c \leftarrow \{ownID\}$,
$chID \leftarrow ownID$,
$clusterhead \leftarrow true$
\While{true}{
	\eIf{clusterhead}{
		\If{ $||c|| \geq 1$}{
			$send(m_{ch})$
		}
	}{
		$send(m_{cm})$
	}

	$wait(T)$
}
\caption{\label{alg1}Agent's main program skeleton}
\end{algorithm}

\paragraph{Request sending and processing}
In addition to the main loop each agent starts a second procedure in parallel,
shown in algorithm\,\ref{alg2}, to send requests to extend the social situation (as mentioned above, this is only possible if the agent acts as a cluster head). The function
\textit{getCandidateForSocialSituation}() returns a list of candidates to
extend the current social situation (based on the aggregated
opinions derived from the other agents' $m_{cm}$ messages).
In case the candidate accepts the request, it becomes the
new cluster head, and the original sender of the request becomes a
cluster member of the extended social situation. In case the addressed agent declines the request since it is
not the cluster head (and forwards the requester to its cluster head), the requester
checks whether it is feasible to merge both social situations (part
of function $check()$ in algorithm\,\ref{alg2}). In case the agent declines because it does not believe in a social situation with
the requesting cluster head, the requesting cluster head marks this request as
failed to prevent flooding the agent with requests.
%
\begin{algorithm}[htb]
 \While{clusterhead}{
	$ID \leftarrow $ getCandidateForSocialSituation()	

\If{$ ID \neq null $}{
		send($m_{req}$)

		$m \leftarrow $ receive($m_{res}$)

		\eIf{ $ m = positive $ }{
			$clusterhead \leftarrow false$

			$chID \leftarrow $ senderID($m$)
		}{ \eIf{ $ m = forward $ $\wedge$ check(senderID($m$),forwardToID($m$))}{
				setAsNextCandidate(forwardToID($m$))
			}
			{
				markRequestAsFailed(senderID($m$))
			}
		}	
	}
{
	waitAPeriodOfTime()
}
}
\caption{\label{alg2}Request sending}
\end{algorithm}

Algorithm\,\ref{alg3} illustrates how an agent replies to a received
$m_{req}$ message: in case the agent is the cluster head of its social
situation, it checks whether an enhanced social situation with the sender of
the message is likely (determined based on the aggregated subjective
logic opinions of the other agents in \textit{check\-For\-Social\-Situation()}),
and sends a positive response message $m_{res}$ in the positive case.
If \textit{check\-For\-Social\-Situation()} returns $false$ (i.e. the existence
of a social situation between the requester and the current cluster
is unlikely), a negative response message $m_{res}$ is sent. If the
agent which received the request is not a cluster head, it replies
with a negative response message pointing the sender to the cluster head
of its social situation.
\begin{algorithm}[htb]
\eIf{clusterhead}{
	\eIf{checkForSocialSituation($m_{req}$)}{
		send($m_{res} \leftarrow (true)$)
	}
	{
		send($m_{res} \leftarrow (false)$)
	}
}
{
	send($m_{res} \leftarrow (false, chID)$)
}
\caption{\label{alg3}Request processing}
\end{algorithm}

\paragraph{Member exclusion}
It is important that the cluster head controls the set of members of
its cluster: if agents leave the cluster, the set of agents within
the cluster propagated in the $m_{ch}$ messages gets adjusted.
%

%
%
\subsection{Making Decisions}
In section\,\ref{sub:Agent-programs}, decision making was shifted to the functions \textit{getCandidateForSocialSituation}() (Algorithm\,\ref{alg2}) and \textit{checkForSocialSituation}() (Algorithm\,\ref{alg3}).
There are two possible ways to aggregate the information
and come to a discrete yes-no-decision:
1) discretize the SL opinions already on agent-level and aggregate yes-no-opinions
later or 2) aggregate the SL opinions and discretize later. 

%
%
%
%
%
%

An early discretization has the benefit of early simplification: A very skeptic agent would only decide in favor
of a social situation if the social situation can be pairwise confirmed
for all potential members. In contrast, a more gullible agent would
vote in support of a social situation even if not all members are
confirmed pairwise.

In contrary, first combining and then discretizing (and thus deciding) offers the
benefit of a more differentiated decision since the confidence of
the individual opinions is considered. Ignoring a single agent with
slight disbelief in favor of four agents with strong opposite belief
seems reasonable, but sparing an agent with a strong belief in favor
of four agents with only slight opposite belief is a less favorable option.

Discretization results in a loss of information and therefore
should be done as late as possible in the decision process. Thus
the latter approach should be preferred. The
individual opinions of all agents within a cluster $c_{i}$ about
being in a social situation with the members of a cluster $c_{j}$
can be combined using the averaging fusion operator $\underline{\oplus}$
for SL opinions \cite{Josang2007b,Groh2011} to build a group opinion $\omega_{c_{i},c_{j}}^{c_{i}}=\underbar{\ensuremath{\oplus}}_{x\in c_{i},y\in c_{j}}\omega_{x,y}$.
After aggregating, the result needs to be discretized using a decision function 
$ f_{\boxplus} $ 
\cite{Groh2011} to allow a mapping to a concrete decision. 
\subsection{Conflict Resolution}
Two agents may send requests to each other exactly
at the same point in time. Thus both would be willing to accept the other's
request. This leads to a conflict as both agents assume to be
the new cluster head. To mitigate this risk the function $checkForSocialSituation()$
in algorithm 3 needs to check whether the requesting agent is an agent
which has been requested to join the social situation previously and which did not reply yet. If this
is the case, the agent with the identifier being closer to the hashed concatenation of both identifiers is the new designated cluster head. 
%
\subsection{Message and Time Complexity}
An upper limit of potential communication partners should exist to ensure scalability \cite{Durfeee2001}. An agent only contacts
other agents if they are either candidates for a social
situation or part of its current social situation (both limited locally). Given $n$ agents
within communication range, the maximum number of messages within a time
interval $t$ is in any case lower than $(2n+1)\cdot(t/f_{min})$
consisting of $n$ request messages $m_{req}$, $n$ response messages
$m_{res}$, and one cluster head or cluster member message per cycle
with frequency $f_{min}$.

An agent stores the current social situation, the identifiers of agents which denied
a request for a social situation, and the identifiers of agents which did not reply
to a request to avoid race conditions. The required storage space depends on the number of agents
involved, with an existing upper bound $n$ since the number of candidates for a social situation is physically limited. All stored information are only kept for a limited time limit. Thus, the required storage space does not increase over time.
\subsection{Optimizations and Variations}
\paragraph{Detach agents from individuals}
\label{sub:Disaggregating-agents-and}
 In a real-world scenario it is more realistic
to assume that some individuals are not represented by an agent and even that 
some agents are not associated with any individual but are fixed
at a specific location. Allowing to split agents and human individuals requires an extension
of the identifier concept: Agents and human individuals which are physically linked (``human related agents'') share an identifier.
Individuals without an agent have an identifier which allows postulating opinions about
social situations for the respective human individual. Agents without
an associated human individual (``opinion providers'') can only provide opinions but cannot be
part of any social situation. This leads to an extended notion of the cluster head message $m_{ch}$
as the message format has to contain information about
real human individuals: $m_{ch}=(ch,c_{a},c_{h})$ with $c_{a}\subseteq c_{h}$
being the set of all agents in the cluster and $c_{h}$ being the
set of all human individuals. Opinion providers are not involved in
the membership process -- they can only send their opinions using $c_{cm}$ messages. A trust concept (like the one proposed by Bamberger et al. \cite{Bamberger2010}) can support the decision which opinions to consider to which degree in building the aggregated group opinion. It is assumed that all agents can identify a human individual as a
single person with a single unique identifier, that agents can recognize
whether humans are equipped with a physically connected agent and
whether an agent is an opinion provider or a real agent. Recognition of human individuals
has to be implemented on a lower architectural level, as it was investigated e.g. in \cite{Cristani2011}.
%
\paragraph{Pseudonyms and identities}
In a Sybil attack a reputation system is subverted by forging identities~\cite{Newsome2004,Douceur2002}.
Any capability of dealing with a certain percentage of malicious acting
nodes can be subverted by forging identities. This can be mitigated
by adding only trustful identities and therefore limiting the number
of identities per node. Trustful identities are usually managed with
the help of a trusted third party acting as a trust anchor for
all agents, handing out identities that are certified through
signatures based on asymmetric cryptography. Approaches to avoid Sybil
attacks without a trusted third party are based on the assumption
that the attacker's resources are limited, see e.g. \cite{Douceur2002}. These approaches successfully prevent mote-class attacks
which use a normal network node. They perform badly in homogeneous network
structures and against laptop-class attacks with potentially unlimited resources.
%
%
%
%
%
\paragraph{Avoiding to request cluster members for social situations}
Sending request messages to an agent which only has member status
within a social situation results in a negative response message. To avoid this unnecessary step every agent should use logged cluster head messages of social situations to gather an
impression about the structure of ongoing social situations and send requests directly to the respective cluster head.
\paragraph{Limitation of opinion weights}
Lower uncertainty can be interpreted as higher weight of an opinion when aggregated using the averaging fusion operator $\underline{\oplus}$. In
order to avoid malicious agents increasing their impact on the system,
minimum levels of uncertainty have to be defined for opinion authors.
\paragraph{Keeping clusters stable}
Instead of dispersing the cluster, a leaving cluster head could nominate a member agent as a replacement cluster head. 
As a first action, the new cluster head would send an $m_{ch}$ message to establish a new cluster with the same members (but without the previous cluster head). 
Possible criteria for selecting a replacement cluster head might be the agent's energy level or the average time within the social situation. This would leverage the observation that agents which move quickly between social situations
would perform poorly as cluster heads (as it is likely that they leave the social
situation quite soon)~\cite{Chatterjee2002,Basagni1999}.

In addition to the extensions listed above, we are aware of a remaining point
of criticism: Whenever an agent postulates an opinion about
two other agents, this opinion is usually not taken into account unless one of the two agents is within the same social
situation as the postulating agent. The protocol does not fully leverage this knowledge as we assume that this has already happened on the analysis level (cf. Fig.\,\ref{fig:architecture}).
\section{Case Study}
\label{sec:evaluation}
%
%
%
%
We simulated the proposed protocol based on numerous datasets of social interactions.
Each agent was equipped with SL opinions about being in a social situation with each other agent.
Without any conceptual restriction to a classifier, we generated the SL opinions with the Gaussian Mixture Models (GMM)-based classifier that was described in \cite{Groh2011b}. 
The classifier uses the relative distances and shoulder angles as input parameters.
We use Rand Index~\cite{Rand1971}, Adjusted Rand Index~\cite{Hubert1985}, and Jaccard Index as a distance measure between the partitions generated by our proposed protocol and the actual social situation clusters.
\subsection{Dataset 1 (DS1)}
The first dataset \cite{Groh2010} contains geometry data of social interaction and was captured with infrared beacons that were fixed on the shoulder of each of the nine interacting participants. Eight cameras tracked the social interaction between the participants. The tracking accuracy was $<1mm$ and $<1\si{\degree}$ (cf. \cite{Groh2010}).
Out of the captured data pairwise relative shoulder angles and pairwise
relative spatial distances have been computed for all participants to reveal a correlation between those parameters and the existence of a social situation.

\subsection{Dataset 2 (DS2)}
The second dataset \cite{cristiani2011} covers two coffee break scenarios
which were used in \cite{Cristani2011}. They were captured by
a single camera.
For our case study, we rely on the positions and orientations of the individuals that were algorithmically computed based on the recorded images (cf.\,\cite{Cristani2011}). We computed the corresponding relative shoulder angles and distances among the participants as input to our simulation.


\subsection{Synthetic Data}
\begin{figure}
\begin{center}
\includegraphics[scale=1]{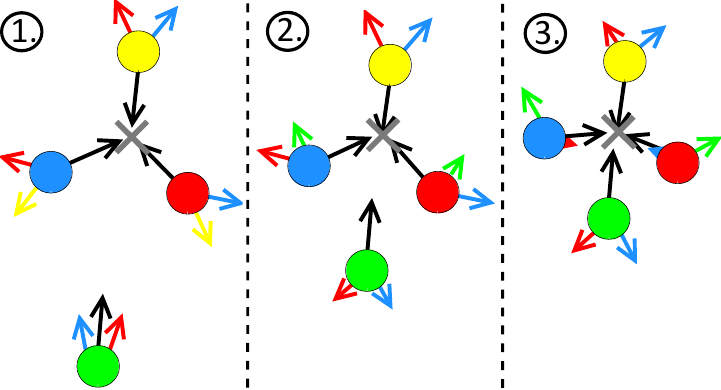}
\end{center}
\caption{\label{fig:grp}Group positioning in enhanced
  ad hoc group simulator SUMI: resting group with 3 members is joined by a fourth member}
\end{figure}

We enhanced the ad hoc group simulator SUMI \cite{Groh2007} 
to support the generation of social situation data. SUMI combines group and node
mobility models that simulate individual Gauss Markov motions of the
nodes. Each node makes a random walk until a random group is formed.
Once all members have arrived, they
wait until the group ends or performs a group motion. 
In comparison to the original version, node speed has been reduced to walking speed, movement speed selection strategy has been replaced by a Markov chain and resting times have been replaced by a force model
for the arrangement of nodes in social situations (inspired by \cite{Brandes2001,Helbing1995,Helbing2002}).
Fig.\,\ref{fig:grp} shows positioning of a resting group with three nodes when a fourth node joins: {``}$\times$'' denotes the group centre,
which attracts all agents. Due to the repelling effect between
the agents, each agent is exposed to the repelling forces of the next two agents. 
The color of the arrows in Fig.\,\ref{fig:grp} indicates the force's origin. 
Shoulder angles are set either with a probabilistic chosen random deviation to the group center or in the moving direction in case of moving groups.
%
Synthetic data has inherent information about the social situations while DS1 and DS2 were annotated manually.

\subsection{Evaluation}

%
%
%
%
%
%

\begin{table}[thb]
\begin{center}

\begin{tabular}{|p{11mm}||p{10mm}|p{11mm}|p{11mm}|p{11mm}|p{11mm}}
\hline 

{Index} &  \multicolumn{2}{c|}{DS1}& {DS2 (1)} & {DS2 (2)}\\\hline
&AVL\cite{Groh2010}&\multicolumn{3}{c|}{Proposed protocol}\\\hline\hline
Rand Index & 0.766 $\pm0.20$ & 0.796 $\pm0.21$  & 0.910 $\pm0.06$ & 0.960 $\pm0.02$\\\hline
ARI &0.529 $\pm0.37$ & 0.571 $\pm0.40$ & 0.093 $\pm0.19$ & 0.116 $\pm0.20$\\\hline
Jaccard Index & $0.67$ & 0.659 $\pm0.31$& 0.074 $\pm0.13$ & 0.082 $\pm0.13$\\\hline
\end{tabular}
\caption{\label{tab:Results-dataset}Key figures for DS1 \& DS2}
\end{center}
\end{table}

Tab.\,\ref{tab:Results-dataset} shows the comparison between real social situations and the social situations as agreed by our protocol, averaged for the complete simulation, for DS1.
We compared the results with the social situations that have been detected by a central clustering approach (AVL) in \cite{Groh2010}.
To ensure comparability, the central clustering approach (AVL) and our proposed protocol used the same SL opinions (which had a correct classification rate of $\sim75\%$ only \cite{Groh2010}). Considering the fact that the benchmark values (AVL) have been achieved using global knowledge, we demonstrated that social situation detection using local knowledge (i.e. our proposed approach) performs at least equivalently.
%
%
%
%
%
%
%
%
%
Both sequences of DS2 lead to poor correct classification rates
when using the GMM of \cite{Groh2010} to generate the SL opinions. 
This can be explained by various reasons:
While the first dataset (DS1) was captured using infrared tracking, the second dataset (DS2) relies on less granular orientation data acquired using image recognition. The foot position is calculated based on the head position and a height
estimation. This results in quite imprecise distances, varying from
16\,$cm$ up to 1.83\,$m$ within a social situation. 
With a recall of $\sim50\%$ only half of the social situations have been recognized. In
\cite{Cristani2011} where the dataset was used originally, both sequences
showed better results for precision and recall (DS2(1): precision
$66\%$, recall $67\%$; DS2(2): precision $85\%$, recall $57\%$).
We conclude that the accuracy of the underlying sensor information is an essential parameter when formulating an appropriate SL opinion in the logical sensors.

\begin{figure}
	\begin{center}
\includegraphics[scale=0.17]{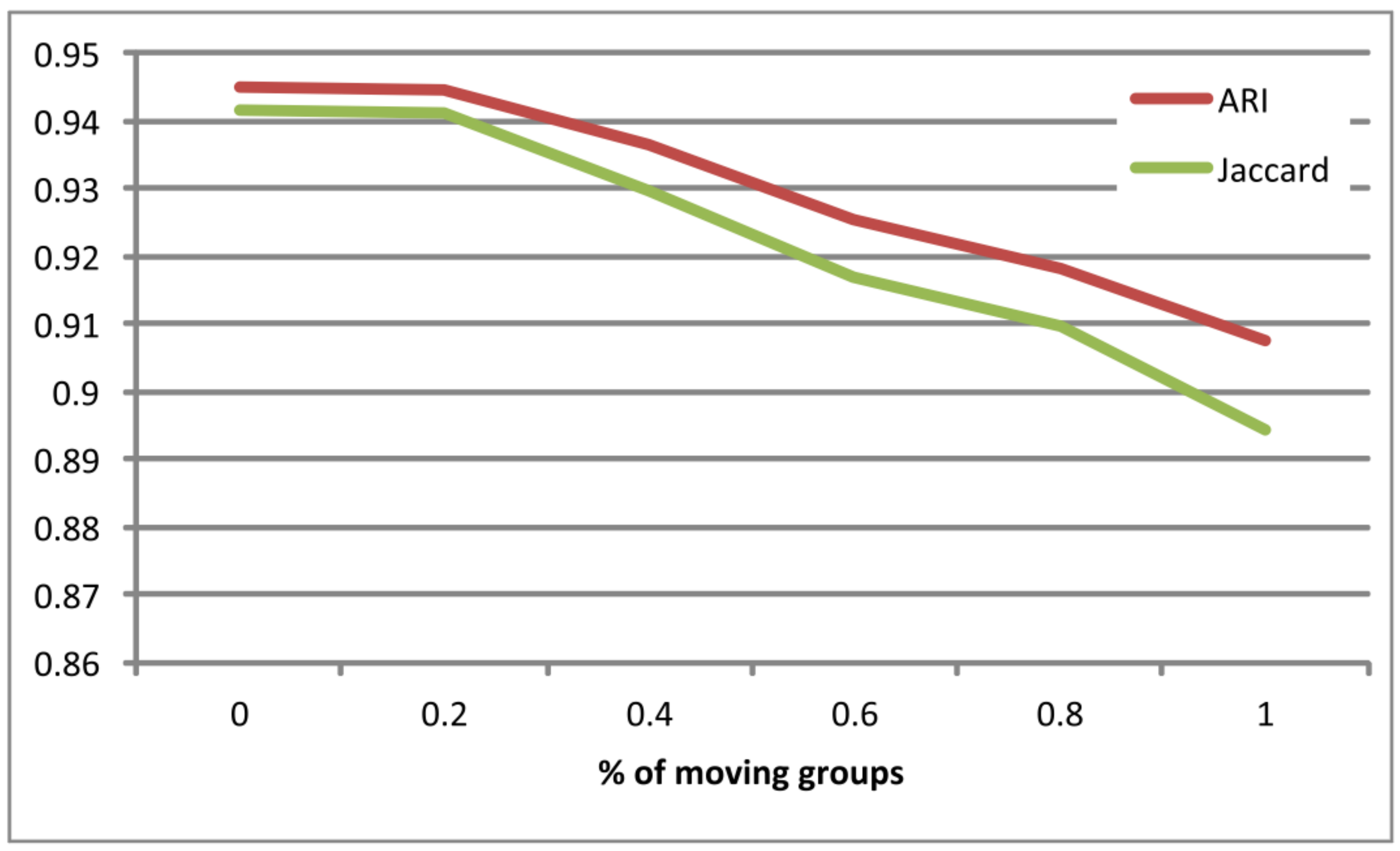}
\end{center}
\caption{\label{fig:mov}Influence of the ratio of moving groups}
\end{figure}
%
\begin{figure}
	\begin{center}
\includegraphics[scale=0.17]{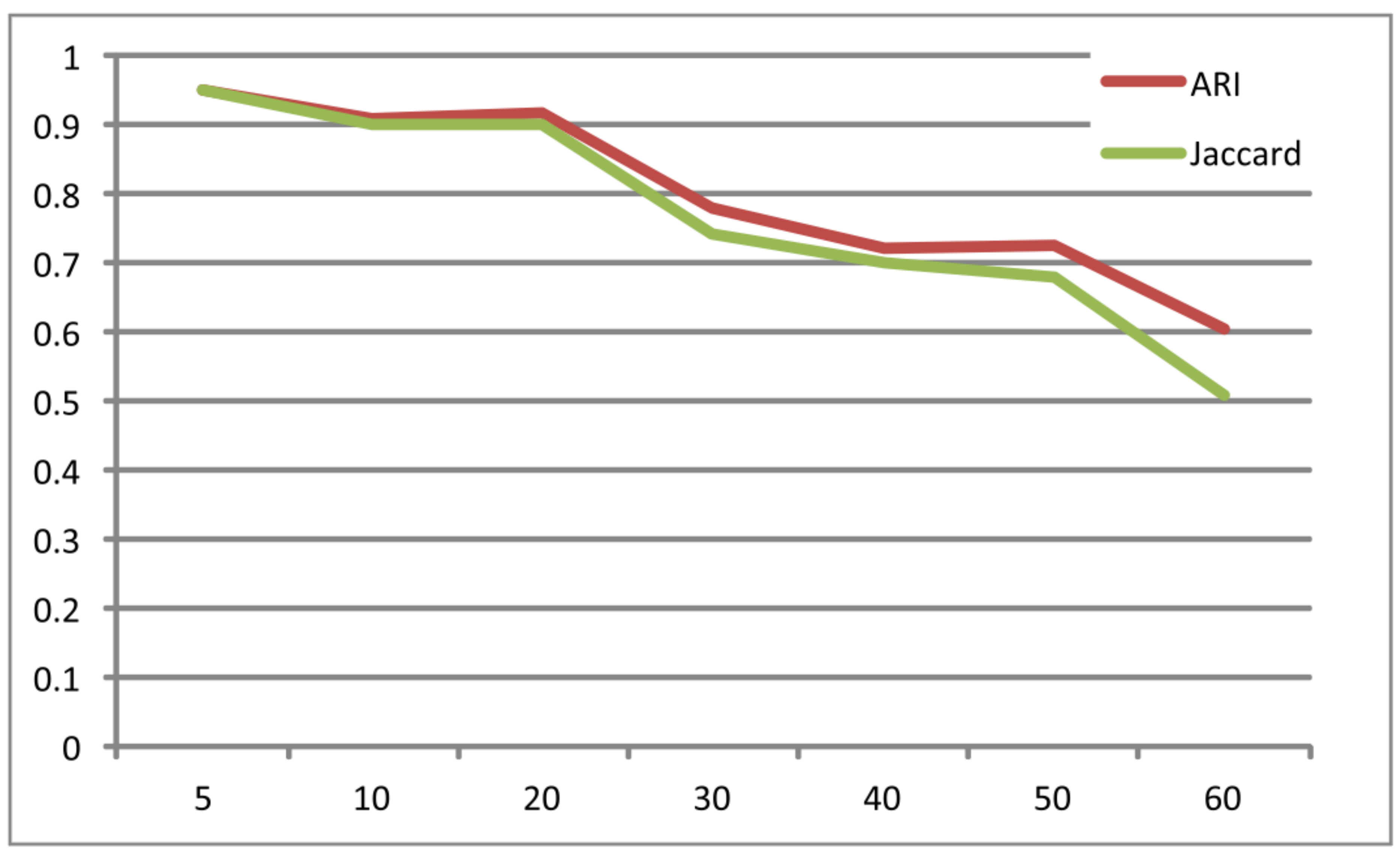}
	\end{center}
\caption{\label{fig:crowded}Influence of the closeness of agents}
\end{figure}

%
\paragraph{Synthetic Data}
Fig.\,\ref{fig:mov} shows Jaccard Index and ARI in relation to the moving group ratio. Rand
Index is not included as it was close to $1$ for all simulations
on synthetic data due to the high number of singletons in the
data. Those singletons are clustered correctly by default and therefore
influence the result positively. In addition to a varying ratio of moving groups, we also investigated
the effect of crowdedness. Fig.\,\ref{fig:crowded} shows Jaccard Index and ARI in relation to the number of simulated agents.
The success of the proposed protocol is highly dependent on the
quality of the underlying SL opinions. Moreover it is
shown that the GMM works well on datasets with low density. If the scenario gets more crowded the number
of false positives (FP) rises significantly.
We recognized a high number of FP (increasing with the number of agents): 2,228 FP for 10, 14,329 FP for 20, and 32,667 FP for 30 agents (simulation of 15 minutes within  $50\times50$ meters). Possible reasons are 1) that the dataset generator does not pay any attention to collision avoidance what causes nodes getting close to each other randomly as if they were in a social situation and 2) that moving groups are not labeled as social situations when two agents are waiting for a third agent to join (however, they are recognized as social situations by the GMM).

\section{Conclusion}
\label{sec:Conclusion}
Our main goal was to model the continuous process of forming, changing, and resolving of social situations. 
Therefore, we introduced a distributed protocol to gain a common understanding of the existing social situations among a group of agents. 
The protocol relies on the aggregation of subjective opinions represented using SL. 
The evaluation demonstrated that it is sufficient to have local knowledge to detect social situations since the proposed protocol performed not worse than traditional cluster techniques requiring full global knowledge of the graph. We demonstrated that the quality of the underlying SL opinions is the limiting factor for the resulting classification quality. Future challenges include scenarios with systems that are able to deal with more inaccurate data to allow real-world applications (e.g. a smartphone in a pocket is unlikely to provide the same data accuracy as a commercial infrared tracking system). In addition, techniques to compare partitions of sets like Rand Index might not be the right measure to compare different clustering results as they punish unimportant delays of social situation detection due to time discretization.
%
%
\bibliographystyle{plain}

\bibliography{lit}  

\begin{thebibliography}{10}

\bibitem{Abbasi2007}
Ameer~Ahmed Abbasi and Mohamed Younis.
\newblock A survey on clustering algorithms for wireless sensor networks.
\newblock {\em Computer Communications}, 30(14-15):2826--2841, 2007.

\bibitem{Agarwal2007}
Ratish Agarwal and Mahesh Motwani.
\newblock Survey of clustering algorithms for manet.
\newblock {\em arXiv:0912.2303}, 2007.

\bibitem{Akbas2013}
M.I. Akbas, R.N. Avula, M.A. Bassiouni, and D.~Turgut.
\newblock Social network generation and friend ranking based on mobile phone
  data.
\newblock {\em Proceedings of the IEEE International Conference on
  Communications (ICC) (Budapest)}, 2013.

\bibitem{Altshuler2012b}
Yaniv Altshuler, Wei Pan, and Alex Pentland.
\newblock Trends prediction using social diffusion models.
\newblock {\em Proceedings of the 2012 International Conference on Social
  Computing, Behavioral-Cultural Modeling, \& Prediction (SBP) (Washington,
  DC)}, 2012.

\bibitem{Bamberger2010}
Walter Bamberger, Josef Schlittenlacher, and Klaus Diepold.
\newblock A trust model for intervehicular communication based on belief
  theory.
\newblock {\em Proceedings of the IEEE International Conference on Social
  Computing (SocialCom) (Minneapolis, MN)}, 2010.

\bibitem{Basagni1999}
Stefano Basagni.
\newblock Distributed clustering for ad hoc networks.
\newblock {\em Proceedings of the International Symposium on Parallel
  Architectures, Algorithms and Networks (Perth/Fremantle, WA)}, 1999.

\bibitem{Brandes2001}
Ulrik Brandes.
\newblock Drawing on physical analogies.
\newblock {\em Lecture Notes in Computer Science}, 2025:71--86, 2001.

\bibitem{Brooks1997}
Richard~R. Brooks and Sundararaja Iyengar.
\newblock {\em Multi-Sensor Fusion: Fundamentals and Applications with
  Software}.
\newblock Prentice Hall PTR, 1997.

\bibitem{Chatterjee2002}
Mainak Chatterjee, Sajal~K. Das, and Damla Turgut.
\newblock Wca: A weighted clustering algorithm for mobile ad hoc networks.
\newblock {\em Cluster Computing}, 5(2):193--204, 2002.

\bibitem{cristiani2011}
M.~Christiani.
\newblock Coffee break dataset, 2011.
\newblock
  \url{http://profs.sci.univr.it/~cristanm/datasets/CoffeeBreak/index.html},
  (checked Feb 2014).

\bibitem{DeGroot1972}
Morris~H. DeGroot.
\newblock Reaching a consensus.
\newblock {\em Journal of the American Statistical Association},
  69(345):118--121, 1974.

\bibitem{Douceur2002}
John~R. Douceur.
\newblock The sybil attack.
\newblock {\em Lecture Notes in Computer Science}, 2429:251--260, 2002.

\bibitem{Durfeee2001}
Edmund~H. Durfee.
\newblock Scaling up agent coordination strategies.
\newblock {\em Computer}, 34(7), 2001.

\bibitem{Groh2007}
Georg Groh.
\newblock Groups and group-instantiations in mobile communities -- detection,
  modeling and applications.
\newblock {\em Proceedings of the International Conference on Weblogs and
  Social Media (Boulder, CO)}, 2007.

\bibitem{Groh2011}
Georg Groh, Christoph Fuchs, and Alexander Lehmann.
\newblock Combining evidence for social situation detection.
\newblock {\em Proceedings of the IEEE International Conference on Social
  Computing (SocialCom) (Boston, MA)}, 2011.

\bibitem{Groh2011b}
Georg Groh and Alexander Lehmann.
\newblock Deducing evidence for social situations from dynamic geometric
  interaction data.
\newblock {\em International Journal of Social Computing and Cyber-Physical
  Systems}, 1(2), 2011.

\bibitem{Groh2010}
Georg Groh, Alexander Lehmann, Jonas Reimers, Marc~Ren{\'e} Frie{\ss}, and
  Loren Schwarz.
\newblock Detecting social situations from interaction geometry.
\newblock {\em Proceedings of the IEEE International Conference on Social
  Computing (SocialCom) (Minneapolis, MN)}, 2010.

\bibitem{Gupta13}
Amarnath Gupta and Ramesh Jain.
\newblock Social life networks: A multimedia problem?
\newblock In {\em Proceedings of the 21st ACM International Conference on
  Multimedia}, MM '13, pages 203--212, New York, NY, USA, 2013. ACM.

\bibitem{Heinzelman2000}
Wendi~Rabiner Heinzelman, Anantha Chandrakasan, and Hari Balakrishnan.
\newblock Energy-efficient communication protocol for wireless microsensor
  networks.
\newblock {\em Proceedings of the Annual Hawaii International Conference on
  System Sciences (HICSS-33)}, 2000.

\bibitem{Helbing2002}
D.~Helbing, I.~Farkas, P.~Moln{\`a}r, and T.~Vicsek.
\newblock Simulation of pedestrian crowds in normal and evacuation situations.
\newblock In M.~Schreckenberg and S.~D. Sharma, editors, {\em Pedestrian and
  Evacuation Dynamics}. Springer, 2002.

\bibitem{Helbing1995}
Dirk Helbing and Peter Molnar.
\newblock Social force model for pedestrian dynamics.
\newblock {\em Physical Review E}, 51(5):4282--4286, 1995.

\bibitem{Hubert1985}
Lawrence Hubert and Phipps Arabie.
\newblock Comparing partitions.
\newblock {\em Journal of Classification}, 2(1):193--218, 1985.

\bibitem{Josang2001}
Audun J{\o}sang.
\newblock A logic for uncertain probabilities.
\newblock {\em International Journal of Uncertainty, Fuzziness and
  Knowledge-Based Systems}, 9(3):279 -- 311, 2001.

\bibitem{Josang2007b}
Audun J{\o}sang.
\newblock Probabilistic logic under uncertainty.
\newblock {\em Proceedings of the Australasian symposium on Theory of computing
  (Ballarat, Victoria)}, 65, 2007.

\bibitem{f-formation2}
Adam Kendon.
\newblock Spacing and orientation in co-present interaction.
\newblock {\em Development of Multimodal Interfaces Active Listening and
  Synchrony}, 5967:1--15, 2010.

\bibitem{Marshall2011}
Paul Marshall, Yvonne Rogers, and Nadia Pantidi.
\newblock Using f-formations to analyse spatial patterns of interaction in
  physical environments.
\newblock {\em ACM Conference on Computer Supported Cooperative Work (CSCW)
  (Hangzhou)}, 2011.

\bibitem{Matic12}
Aleksandar Matic, Venet Osmani, Alban Maxhuni, and Oscar Mayora.
\newblock Multi-modal mobile sensing of social interactions.
\newblock In {\em PervasiveHealth}, pages 105--114, 2012.

\bibitem{Newsome2004}
James Newsome, Elaine Shi, Dawn Song, and Adrian Perrig.
\newblock The sybil attack in sensor networks: Analysis \& defenses.
\newblock In {\em Proceedings of the 3rd International Symposium on Information
  Processing in Sensor Networks}, IPSN '04, pages 259--268, New York, NY, USA,
  2004. ACM.

\bibitem{Olfati2007}
Reza Olfati-Saber, J.~Alex Fax, and Richard~M. Murray.
\newblock Consensus and cooperation in networked multi-agent systems.
\newblock {\em Proceedings of the IEEE}, 95(1), 2007.

\bibitem{Cristani2011}
Giulia Paggetti, Andrea Fossati, Diego Tosato, Alessio~Del Bue, Gloria Menegaz,
  Marco Cristani, Loris Bazzani, and Vittorio Murino.
\newblock Social interaction discovery by statistical analysis of f-formations.
\newblock {\em Proceedings of the British Machine Vision Conference (Dundee)},
  2011.

\bibitem{Pan2011}
Wei Pan, Nadav Aharony, and Alex Pentland.
\newblock Fortune monitor or fortune teller: understanding the connection
  between interaction patterns and financial status.
\newblock {\em Proceedings of the IEEE International Conference on Social
  Computing (SocialCom) (Boston, MA)}, 2011.

\bibitem{Rand1971}
William~M. Rand.
\newblock Objective criteria for the evaluation of clustering methods.
\newblock {\em Journal of the American Statistical Association},
  66(336):846--850, 1971.

\bibitem{Shafer1971}
Glenn Shafer.
\newblock {\em A Mathematical Theory of Evidence}.
\newblock Princeton University Press, 1971.

\bibitem{Shayeb2011}
I.~G. Shayeb, A.~H. Hussein, and A.~B. Nasoura.
\newblock A survey of clustering schemes for mobile ad-hoc network (manet).
\newblock {\em Americal Journal of Scientific Research}, (2):135--151, 2011.

\bibitem{Vinciarelli2009}
Alessandro Vinciarelli, Maja Pantic, and Herv{\'e} Bourlard.
\newblock Social signal processing: Survey of an emerging domain.
\newblock {\em Image and Vision Computing}, 27(12):1743--1759, 2009.

\bibitem{Wooldrige2002}
M.~Wooldrige.
\newblock {\em An introduction to multiagent systems}.
\newblock John Wiley \& Sons, 2002.

\bibitem{Younis2004}
Ossama Younis and Sonia Fahmy.
\newblock Distributed clustering in ad-hoc sensor networks: A hybrid,
  energy-efficient approach.
\newblock {\em Proceedings of the Annual Joint Conference of the IEEE Computer
  and Communications Societies (INFOCOM) (Hong Kong)}, 2004.

\bibitem{Yu2005}
Jane~Y. Yu and Peter H.~J. Chong.
\newblock A survey of clustering schemes for mobile ad hoc networks.
\newblock {\em IEEE Communications Surveys \& Tutorials}, 7(1), 2005.

\end{thebibliography}
\end{document}